# A LSTM-Transformer Model for pulsation control of pVADs

Chaoran E[1], Chenghan Chen[1], Yuyang Shi[1], Haiyun Wang[1], Peixin Hua[1], Xiwen Zhang[1,*]

1 Department of Engineering Mechanics, School of Aerospace Engineering, Tsinghua University, Beijing 100084, China

* Correspondence:
Xiwen Zhang
zhangxiw@tsinghua.edu.cn

Keywords: Percutaneous ventricular assist device, pulsatile control method, NPQ Model, LSTM-Transformer Model, High robustness, Hydraulic and animal experiments validation

## Abstract

**Research background and purpose:** For high-risk percutaneous coronary intervention (PCI) and cardiogenic shock(CS) patients, a percutaneous ventricular assist device (pVAD) is one of the most effective mechanical circulation support devices(MCS). However, currently most pVADs use a constant flow infusion method. Long-term use of a constant flow infusion device may increase the risk of adverse outcomes. Consequently, investigating the pulsatile function of a pVAD has significant clinical implications.

**Methods:** A method of the pulsation for a pVAD is proposed (AP-pVAD Model). AP-pVAD Model consists of two parts: NPQ Model and LSTM-Transformer Model. (1)The NPQ Model determines the mathematical relationship between motor speed, pressure, and flow rate for the pVAD. (2)The Attention module of Transformer neural network is integrated into the LSTM neural network to form the new LSTM-Transformer Model to predict the pulsation time characteristic points for adjusting the motor speed of the pVAD.

**Results:** The AP-pVAD Model is validated in three hydraulic experiments and an animal experiment. (1)The pressure provided by pVAD calculated with the NPQ Model has a maximum error of only 2.15 mmHg compared to the expected values. (2)The pulsation time characteristic points predicted by the LSTM-Transformer Model shows a maximum prediction error of 1.78ms, which is significantly lower than other methods. (3)The in-vivo test of pVAD in animal experiment has significant improvements in aortic pressure. Animals survive for over 27 hours after the initiation of pVAD operation.

**Conclusion:** (1)For a given pVAD, motor speed has a linear relationship with pressure and a quadratic relationship with flow. (2)Deep learning can be used to predict pulsation characteristic time points, with the LSTM-Transformer Model demonstrating minimal prediction error and better robust performance under conditions of limited dataset sizes, elevated noise levels, and diverse hyperparameter combinations, demonstrating its feasibility and effectiveness.

# 1.Introduction

Heart failure is one of the leading causes of mortality worldwide, affecting over 26 million individuals [1]. Over the past half century, Left Ventricular Assist Device (LVAD) technology has evolved into a recognized therapeutic option for heart failure[2]. A common manifestation of heart failure is cardiogenic shock(CS)[3]. LVADs are widely utilized as an effective therapeutic measures for patients with CS[4].

Currently, there is a diverse variety of LVADs available. These devices can be classified based on their mode of assistance into the pulsatile and the continuous flow types[5]. The pulsatile mode can provide perfusion that mimics the rhythmic pressure and flow rate variations of the human heart[6]. The continuous flow mode, on the other hand, provides a constant pressure and flow rate. At present, the majority of LVADs are of the continuous flow type[5]. However, prolonged use of continuous flow LVADs may lead to adverse symptoms such as gastrointestinal bleeding, pulmonary hypertension, and ventricular suction[7]. Therefore, the study of artificial pulsation in pVADs is both necessary and clinically significant.

Until when two critical issues are addressed would pulsatile blood flow be realized: 1. Establishing the relationships between motor speed, pressure, and flow rate to ensure that the pVAD meets the clinical requirements for pressure and flow values[8, 9]. 2. Determining the precise timing for motor speed adjustment. The transmission, processing, and response of the signals between the motor and the sensor may introduce delays, thus predicting interval time in advance is necessary. In clinical surgeries, each patient has a distinct heart rate, which means that the pulsation characteristic time points of each cardiac cycle vary from one individual to another[10]. Furthermore, the pulsation characteristic time points of a patient may undergo significant alterations due to various unexpected intraoperative events[11]. Therefore, the approach of providing pulsatile blood flow based on fixed time intervals is unfeasible. A time series prediction model is required to forecast the pulsation characteristic time points for several future cardiac cycles.

Traditional time series forecasting methods, such as ARIMA[12, 13] and Weighted Moving Average[14], require specific assumptions about the data. However, experimental and clinical data are commonly complex, which are not suitable for these methods. Moreover, these methods fail to capture complex nonlinear long-term dependencies[15] in time-series data. This adversely affects their predictive accuracy in practical applications[16].

With the rapid advancements in deep learning theory and practice, numerous methods for time series forecasting utilizing deep learning[17] and big data[18] have been proposed. Compared to traditional time series forecasting methods, these approaches have demonstrated significant improvements in prediction performance. Early methods for time series forecasting based on deep learning and big data utilized models such as RNN[19], GRU[20], and LSTM[21], achieving better predictive results than traditional methods. However, these models are often limited to using shorter time series data[22] as input, whereas the information contained in short-term time series is insufficient to fully capture the rich temporal features present in historical data[23]. Additionally, there is an alternative way to put each time step of long time series directly into predicting models, such as the Transformer[24] which excels at processing long sequences. However, this forecasting method tends to have lower processing efficiency[25].The LSTM-Transformer model is capable of addressing this issue. The principle of the model is as follows: LSTM uses a gating mechanism[26] to effectively capture long-term dependencies in sequences, preventing gradient vanishing or explosion, thereby alleviating the long-term dependency[27] problem to some extent. However, due to its inherently recursive structure[28], it cannot truly resolve the long-term dependency issue. In contrast, the Transformer model based on self-attention mechanisms[29] employs a parallel attention mechanism that captures sequence dependencies comprehensively through self-attention[30] and multi-head attention[31], enabling it to consider all positions in the input sequence simultaneously. This allows it to extract more distant temporal features[32] and better understand the relationships between previous and subsequent data points[33]. The advantage of the LSTM-Transformer model over traditional LSTM models is its ability to evade the long-term dependency problem[34] while achieving efficient parallel computation[35].The LSTM-Transformer model is developed for predicting pulsation characteristic time points in this work. The LSTM-Transformer model has structural advantages[36] over traditional time series forecasting methods.

In this work, an artificial pulsatile model for pVADs (AP-pVAD Model) is proposed. It consists of two sub-models: 1. A mathematical model describing the relationship between motor speed, blood pump pressure, and blood pump flow (NPQ Model). 2. The new model called LSTM-Transformer Model, which is trained using an animal experimental dataset, allowing it to predict the pulsation characteristic time points that adjust motor speed. The motor controller adjusts the motor speed by integrating the results from both models to provide pulsatile blood flow. Finally, these models are applied in hydraulic and animal experiments to validate their feasibility and effectiveness. As in Fig1, this study conduct several hydraulic experiments and animal experiments, with the datasets from these experiments used as solution data and training data for the NPQ Model and LSTM-Transformer Model respectively. This part of the experiments is mentioned as 'Previous Experiments'. After the NPQ Model and LSTM-Transformer Model are established, both models are deployed to the motor controller of the pVAD, followed by three hydraulic experiments and one animal experiment for validation. This part of the experiments is referred to as 'New Experiments'.

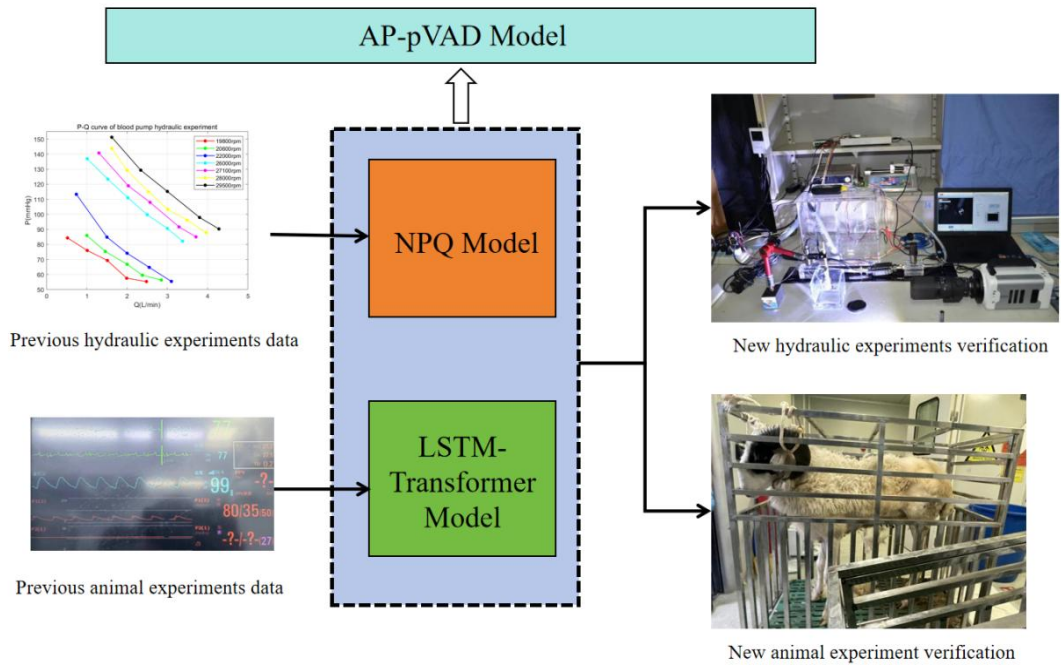

Fig. 1 Framework for studying pVADs artificial pulsatile function

## 2.Methods

## 2.1 Experimental subjects

The blood pump used in this study is a 6 mm diameter micro-axial flow pVAD developed by Tsinghua University[8]. As in Fig2, this pVAD consists of components such as an impeller, a diffuser, a micro motor, and a shroud.

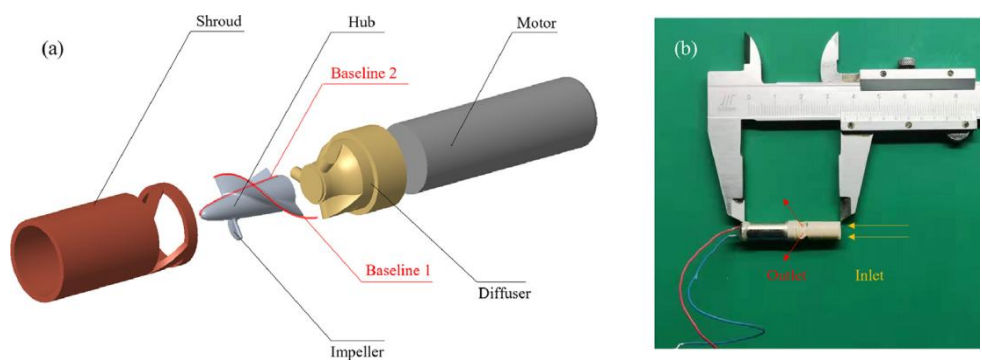

Fig.2 (a) a 3D modeling of the main components of the pVAD (b) a prototype used for experimental testing

### 2.2.1 Establishment of the NPQ Model

To investigate the relationships among the rotational speed, pressure, and flow of the pVAD, a mathematical model for motor speed, pressure, and flow (NPQ Model) is proposed. The specific form of the model is hypothesized as Equation(1):

$$N(t) = a[P(t)]^\alpha + b[Q(t)]^\beta + C \tag{1}$$

where $N(t)$ represents the motor speed of the blood pump, $P(t)$ represents the pressure of the blood pump, and $Q(t)$ indicates the output flow rate of the blood pump. are both unknown parameters.

Based on the derivation from Bernoulli's equation and other relevant theoretical foundations, the values of $a, b, \alpha, \beta$ can be identified. This reduces the number of parameters to solve to two, significantly increasing computational efficiency.

Before derivation, the following assumptions are made:
1. It is assumed that the fluid within the blood pump is considered to be a steady, incompressible ideal fluid[37].
2. It is assumed that the blood moves along the streamline in unit volume due to the elongated internal space of the blood pump's helical tube[8].
3. It is assumed that the work done by the blood pump on a unit volume of blood is assumed to have no energy losses[38].

The process of deriving the two power parameters in the mathematical model is as follows: Because the size of the pVAD is small, the height difference between the pump inlet and outlet is negligible; thus, the fluid conditions at the inlet and outlet of the pVAD can be approximately considered as inviscid. In this inviscid case, the Bernoulli equation is applied at both the pump inlet and outlet as Equation(2).

$$\begin{cases} \dfrac{p_1}{\rho} + \dfrac{v_1^2}{2} = C_1 \\ \dfrac{p_2}{\rho} + \dfrac{v_2^2}{2} = C_2 \end{cases} \tag{2}$$

where $\rho$ represents the density of human blood, $p_1$ represents the pressure of the blood at the pump inlet, $p_2$ represents the pressure of the blood at the pump outlet, $v_1$ represents the flow velocity of the blood at the pump inlet, $v_2$ represents the flow velocity of the blood at the pump outlet, $C_1$ and $C_2$ are constants.

Under the condition that gravitational potential energy does not perform work, the change in the energy of the blood fluid within the blood pump is equal to the work done on it by the pump, which can be expressed as Equation(3):

$$C_2 - C_1 = \frac{N \cdot T \cdot (t_2 - t_1)}{\rho \cdot V} \tag{3}$$

where $N$ represents the pump motor speed, $T$ represents the pump motor torque, $t_1$ represents the time taken for the blood to flow through the inlet, $t_2$ represents the time taken for the blood to flow through the outlet, and ( e ) represents the volume of blood within the pump. The volume of blood within the pump can be expressed as Equation(4):

$$V = S \cdot \int_{t_1}^{t_2} v(t)dt \tag{4}$$

where $S$ represents the cross-sectional area of the blood flow within the pump. Organizing the above equation as Equation(5):

$$\frac{\Delta P}{\rho} + \frac{1}{2}\Delta(v^2) = \frac{N \cdot T \cdot (t_2 - t_1)}{\rho \cdot S \cdot \int_{t_1}^{t_2} v(t)dt} \tag{5}$$

where $\Delta P$ represents the pressure difference across the pump, and $\Delta(v^2)$ represents the difference in the squares of the fluid velocities at both ends of the pump. By applying the Mean Value Theorem for Integrals, the integral term can be manipulated as Equation(6):

$$\frac{\Delta P}{\rho} + \frac{1}{2}\Delta(v^2) = \frac{N \cdot T \cdot (t_2 - t_1)}{\rho \cdot S \cdot v(\xi)(t_2 - t_1)} \tag{6}$$

The above expression has been simplified as Equation(7):

$$\frac{\Delta P}{\rho} + \frac{1}{2}\Delta(v^2) = \frac{N \cdot T}{\rho \cdot S \cdot v(\xi)} \tag{7}$$

The objective of this study is to investigate the relationship between motor speed, pressure, and flow rate. Consequently, motor speed is regarded as the independent variable in the formulation of relevant equations as Equation(8):

$$N = \frac{\Delta P \cdot S \cdot v(\xi)}{T} + \frac{\Delta(v^2) \cdot S \cdot v(\xi) \cdot \rho}{2T} \tag{8}$$

According to the definition of flow rate, it is the volume of fluid that passes through a specific cross-section in a unit of time as Equation(9):

$$Q = S \cdot v(\xi) \tag{9}$$

Derive the expression for motor speed as Equation(10):

$$N = \frac{\Delta P \cdot Q}{T} + \frac{\Delta(v^2) \cdot \rho \cdot Q}{2T} \tag{10}$$

Compared to the research objective, the above equation includes factors such as motor speed, pressure, and flow rate. However, it introduces an additional term $\Delta(v^2)$ that requires further consideration. Furthermore, since the blood within the pump is an incompressible steady flow fluid, the flow rates at the pump inlet and outlet remain constant according to the law of conservation of mass and the continuity equation. This leads to $Q_1 = Q_2 = Q$, where $Q$ is a fixed constant as Equation(11):

$$\Delta(v^2) = v_2^2 - v_1^2 = \frac{Q_2^2}{S_2^2} - \frac{Q_1^2}{S_1^2} = Q^2(\frac{S_1^2 - S_2^2}{S_1^2 \cdot S_2^2}) \tag{11}$$

The inlet and outlet cross-sectional areas of a pump are fixed values for each specific pump. Considering that different pump designs may vary, specific numerical values are not provided here. For a given blood pump as Equation(12):

$$S_2 = k \cdot S_1 \tag{12}$$

where $k$ is a fixed constant. Substituting into the above equation as Equation(13):

$$\Delta(v^2) = Q^2(\frac{S_1^2 - S_2^2}{S_1^2 \cdot S_2^2}) = Q^2(\frac{1 - k^2}{S_2^2}) = v_2^2(1 - k^2) \tag{13}$$

It can be seen from the above equation that $\Delta(v^2)$ is directly proportional to the square of the blood flow velocity $v_2^2$ at the pump outlet. Therefore, the expression for the rotational speed at the pump outlet is given by as Equation(14):

$$N = \frac{1}{T} \cdot \Delta P \cdot Q + \frac{h}{2T} \cdot \rho \cdot v^2 \cdot Q \tag{14}$$

where $h$ is a specific constant. Additionally, because of $v = \frac{Q}{S}$, the above equation can be

rewritten as Equation(15):
$$N = \frac{1}{T} \cdot \Delta P \cdot Q + \frac{h}{2T} \cdot \rho \cdot \frac{Q^3}{S^2} \tag{15}$$
where $h$ is a constant. The blood flow density $\rho$ in the human body is a fixed value. The cross-sectional area $S$ at the pump outlet is also a fixed value. According to the reference[39], the torque of the blood pump motor is directly proportional to the flow rate as Equation(16):
$$T = m \cdot Q + n \tag{16}$$
where $T$ represents the motor torque, while $m$ and $n$ are constants. The above equation can be expressed as Equation(17):
$$N = \frac{1}{mQ + n} \cdot \Delta P \cdot Q + \frac{h \cdot \rho}{2S^2} \cdot \frac{1}{mQ + n} \cdot Q^3 \tag{17}$$
To simplify the mathematical relationship and avoid the coupling terms resulting from the multiplication of $\Delta P$ and $Q$, $T = m \cdot Q$ is approximated. Additionally, to minimize the errors introduced by this approximation, we introduce correction parameters $a$ and $b$, as well as a constant term $C$ [40]. Therefore, the above equation is approximately equal to as Equation(18):
$$N \approx \frac{1}{mQ} \cdot \Delta P \cdot Q + \frac{h \cdot \rho}{2S^2} \cdot \frac{1}{mQ} \cdot Q^3 + C = a \cdot \Delta P + b \cdot Q^2 + C \tag{18}$$
where $a, b, C$ are constants. Let the motor speed be represented as a function of time, denoted by $N(t)$; the pressure difference across the pump be represented as a function of time, denoted by $P(t)$; and the blood flow rate be represented as a function of time, denoted by $Q(t)$. Thus, the above equation can be expressed as Equation(19):
$$N(t) = a \cdot P(t) + b \cdot Q^2(t) + C \tag{19}$$
In conclusion, the mathematical model for the motor speed, pressure, and flow rate of the pVAD is derived. The motor speed is linearly related to the square of the pressure and flow rate. By substituting the hydraulic experimental data of the pVAD into this model, and can be determined.

### 2.2.2 The determination of parameters for the NPQ model

Hydraulic experiments on the pVAD are conducted under various experimental conditions (different rotational speeds, pressures, and flow rates). For each condition, several experiments are performed, and the average value of these three experiments is recorded as the experimental data for that condition, which is then plotted on the P-Q curve below, as in Fig3.

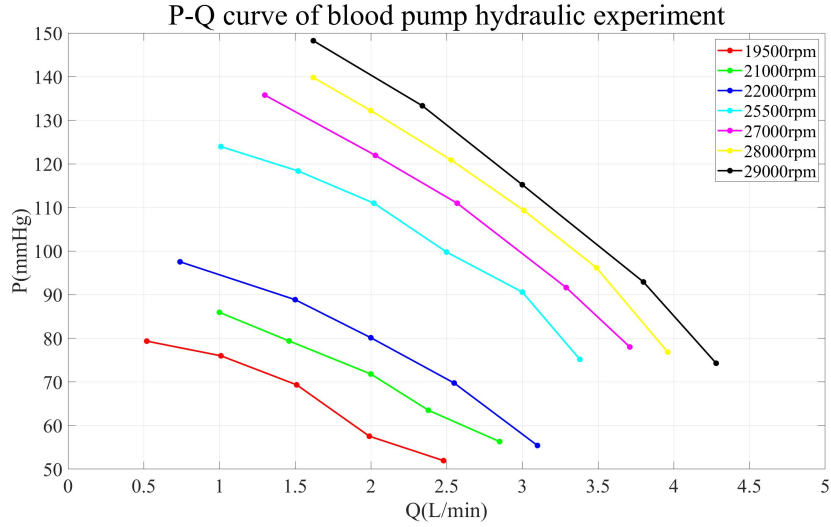

Fig.3 Hydraulic Experiment P-Q Curve of the pVAD

From the P-Q curve, it is evident that the pVAD used in this experiment can provide pressures ranging from 55 mmHg to 150 mmHg with flow rates between 0.5 L/min and 4.3 L/min, and the greater the flow provided by the pump, the smaller the pressure difference across the pump. Furthermore, during high-risk PCI, the pVAD must provide a minimum flow rate of 2.5 L/min[40]. The figure contains a substantial number of operating points that fall within this range. Additionally, higher motor speeds increase the likelihood of the adverse events such as hemolysis and thrombosis[8]. In this experiment, the motor speed of the pVAD achieved a flow rate of 4 L/min at speeds less than 29,000 r/min which provides a significant redundancy compared to the clinical requirement of 2.5 L/min.

By substituting the dataset from the hydraulic experiments of the pVAD into the aforementioned mathematical model, the coefficients $a$ and $b$ are solved, resulting in the following conclusion:

$$N(t) = 117.4P(t) + 551Q^2(t) + 10270 \qquad (6)$$

where $N(t)$ represents the motor speed, measured in $r/min$. $P(t)$ represents the pressure provided by the pump, measured in $mmHg$. $Q(t)$ indicates the flow rate delivered by the pump, measured in $L/min$. All coefficients in the model are accompanied by units. The unit for coefficient 117.4 is $r/(mmHg \cdot min)$, the unit for coefficient 551 is $(r \cdot min)/L^2$, and the unit for coefficient 10270 is $r/min$.

## 2.3 Establishment of the LSTM-Transformer Model

To study the pulsatile performance of the pVAD, the rise points of AOP in each cardiac cycle from long-term continuous time series data are collected during animal experiments[5]. The frequency of data collection in the animal experiments is 150 Hz. To simplify the representation of pulsation time characteristic points, the time interval between each current pulsation time feature point and the previous one is calculated and recorded. Using 1000 milliseconds as a baseline, points that precede it by $m$ milliseconds are marked as $+m$ in the characteristic pulsation time points dataset, whereas points that lag behind it by $n$ milliseconds are marked as $-n$ in the dataset[8].

For example, if the time interval between the current pulsation time feature point and the previous one is 880 ms, this characteristic pulsation time point is recorded as +120.

The LSTM-Transformer Model is established to predict future characteristic pulsation time characteristic points. The constructed LSTM-Transformer framework is presented in Table.1 :

Table.1 The framework and parameters of the LSTM-Transformer model

| Layer（type） | Output Shape | Param # |
| --- | --- | --- |
| input_1(InputLayer) | (None,6,15) | 0 |
| LSTM0(LSTM) | (None,6,8) | 768 |
| multi_head_attention(MultiHeadAttention) | (None,6,8) | 8968 |
| layer_normalization(LayerNormalization) | (None,6,8) | 16 |
| flatten(Flatten) | (None,48) | 0 |
| dense(Dense) | (None,1) | 49 |

The LSTM-Transformer model consists of the following modules:

**LSTM Layer:** This layer leverages the powerful time feature learning capabilities of LSTM neural networks. It efficiently mitigates the issues of gradient vanishing and exploding, which significantly enhances the model's ability to predict time series data.

**Multi-Head Attention Layer:** This layer comprehensively captures sequence dependencies and considers all positions within the input sequence simultaneously. It is capable of extracting distant temporal features, thereby improving the understanding of relationships between preceding and subsequent data points and avoiding long-term dependency problems[16].

**Layer Normalization:** This component helps alleviate the problems of gradient vanishing and explosion during the training process, making the model easier to train.

**Flatten Layer:** This layer converts multi-dimensional inputs into one-dimensional arrays, facilitating the transition from convolutional layers to dense layers.

**Dense Layer:** In this layer, the features extracted by previous layers undergo nonlinear transformations, allowing the model to uncover associations between these features before mapping them to the output space.

An overall schematic diagram of the LSTM-Transformer model is shown in Fig.4:

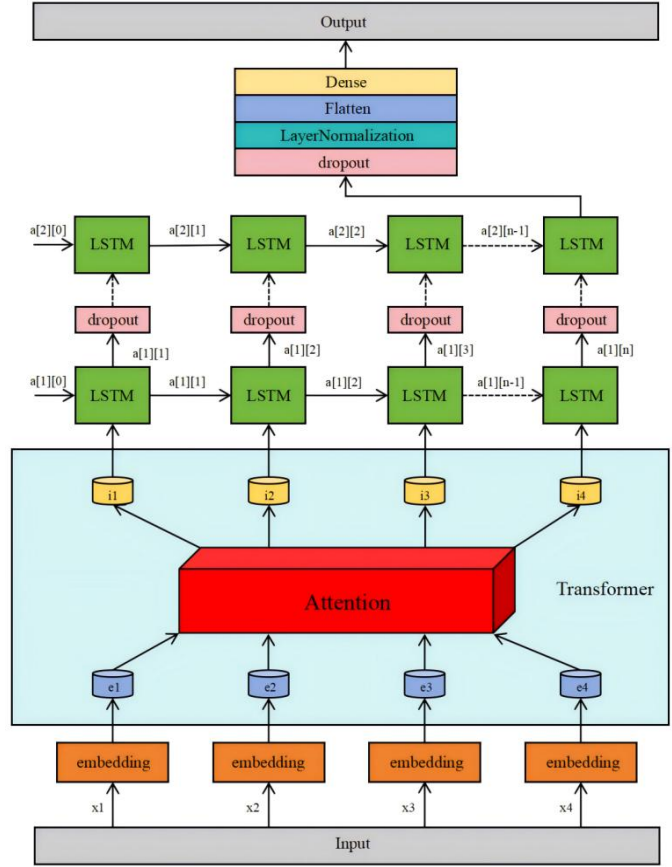

Fig.4 LSTM-Transformer Model Schematic. The input data is transformed into a format that is easily understood by the Attention layer through the embedding layer. The Attention layer learns comprehensive features of the data and assigns different weights to the features of each time series data point. The processed results from the Attention layer are then fed into the LSTM layer and Dropout layer for further learning, ultimately achieving accurate predictions of the time series data.

To evaluate the performance of the LSTM-Transformer model, the following metric is employed:

**Mean Absolute Error (MAE):** MAE measures the average absolute deviation between predicted values and actual values, providing a clear indication of the real errors in predictions.

$$MAE = \frac{1}{n}\sum_{t=1}^{n}|\hat{y}_t - y_t| \qquad (7)$$

**Mean Squared Error (MSE):** MSE is defined as the average of the squared differences between the predicted values and the actual values. It is a useful metric for assessing the variability of a model's predictions and provides insight into how well the model captures the underlying trend of the data. MSE penalizes larger errors more severely than smaller ones, which can be particularly useful when outliers are present in the dataset.

$$MSE = \frac{1}{n}\sum_{t=1}^{n}(\hat{y}_t - y_t)^2 \qquad (8)$$

where $\hat{y}_t$ represents the predicted values, $y_t$ represents the actual values, and $n$ is the number of samples. Generally, the smaller the values of MAE and MSE, the less the error between the predicted values and the actual values. This indicates improved predictive performance of the model. Therefore, lower MAE and MSE values signify a better-fitting model, capable of accurately predicting the future data.

## 2.4 The construction of the pVAD hydraulic experimental platform

To validate the feasibility and effectiveness of the NPQ Model and LSTM-Transformer Model in practical applications, we applied them to both hydraulic experiment and animal experiments of pVADs. A hydraulic experimental platform for the pVAD is constructed, as illustrated in Fig.5:

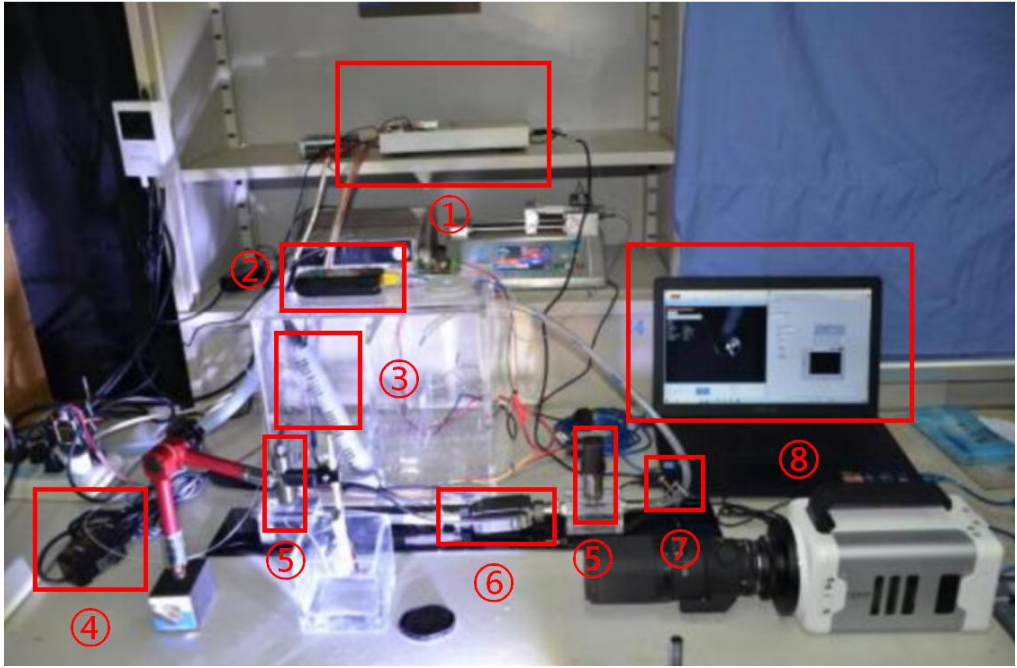

Fig.5 Hydraulic Experimental Platform for the pVAD

**Principle of Hydraulic Experiments:** The tank simulates the human environment and is divided into two parts by an overflow partition. The pVAD pumps water from the left tank to the right tank. The height of the overflow partition is lower than that of the tank walls. Once the water in the right tank accumulates to a certain level, it overflows back to the left tank, thereby creating a water circulation system. The solution used in the hydraulic experiment consists of deionized water, alcohol, and glycerol in a ratio of 1:1:1[8]. The pVAD primarily comprises an impeller and a micro motor, with the micro motor driving the rotation of the impeller. After the assembly of the impeller and micro motor is complete, they are enclosed within a protective casing.

**Components of Hydraulic Experiments:**
1.**Adjustable Constant Voltage Source:** This allows for real-time adjustment of input voltage.

**2.Temperature Sensor:** Real-time measurement of the motor surface temperature.
**3.Constant Temperature Heat Source:** Located on the left side of the tank, it is set to maintain a constant temperature of 37°C to closely simulate the internal thermal environment.
**4.Flow Meter:** This is an ultrasonic flow meter.
**5.Pressure Sensor:** Measurement of pressure values at the pump's inlet and outlet.
**6.Pump:** The core component of the pVAD.
**7.Shut-off Valve:** The function of the shut-off valve is to regulate the flow within the waterway, enabling the simulation of various operational conditions.
**8.Computer:** Data Visualization.

## 2.5 The surgical procedure for pVAD animal experiments

Based on the completion of the aforementioned hydraulic experiments, animal experiments were conducted. The animal subjects used in the experiments are healthy male sheep. After establishing a blood circulation pathway within the ovine body using artificial vessels and other equipment, the pVAD is inserted through the descending aorta. Invasive pressure monitoring devices, such as pressure guide wires, are used to collect pressure signal data from the ascending aorta, which is then input into the NPQ Model and LSTM-Transformer Model. The models calculate the results and output them to the motor controller. The controller adjusts the motor speed to achieve pulsatile blood flow.

# 3. Result

## 3.1 Comparison of the performance of the training and testing sets for three time series prediction models

The dataset used in this study consists of the time series data of characteristic pulsation time points from animal experiments, totaling 12,251 data points. The computational environment for the LSTM-Transformer model is an NVIDIA GeForce RTX 4060 Laptop GPU. The LSTM-Transformer model is implemented using Python and the TensorFlow framework. To achieve faster speed, higher accuracy, and more robust performance in predictions, this study constructs two additional commonly used time series prediction models—LSTM and GRU—and conducts a multidimensional comparison with the LSTM-Transformer model. The parameters for the LSTM and GRU models, such as iteration number, learning rate, and batch size, are kept consistent with those of the LSTM-Transformer model. The iteration count for all three models is set to 100. Iteration terminates when the error threshold is less than $10^{-6}$. The pulsation time characteristic points are input into the LSTM-Transformer, GRU, and LSTM models, using MSE as the loss function to generate the loss curves for the training processes of the three models.

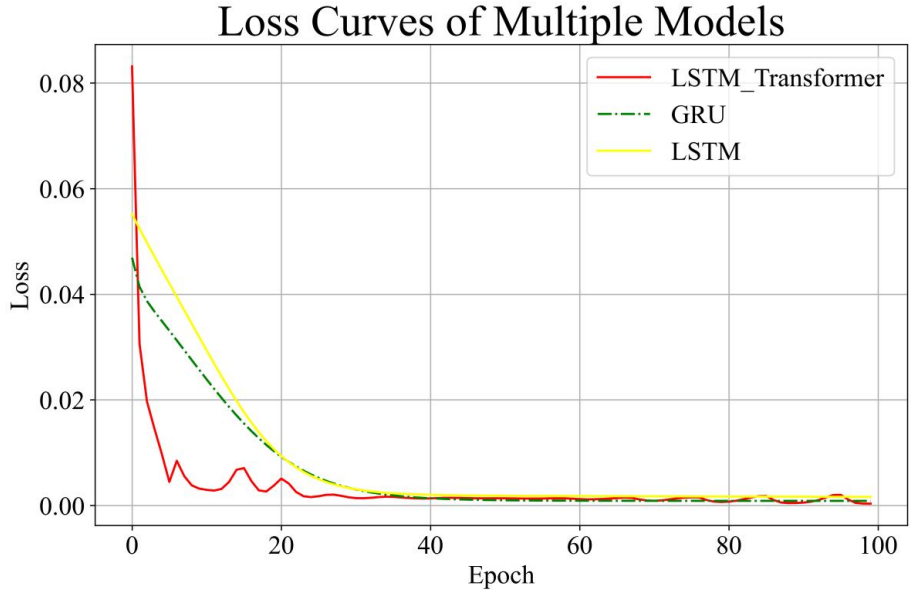

Fig.6 The loss function curves during the training processes of three models. The red line represents the variation in the loss function value of the LSTM-Transformer Model with respect to the number of iterations, the green line represents the variation in the loss function value of the GRU Model, and the yellow line represents the variation in the loss function value of the LSTM Model with respect to the number of iterations.

As indicated in Fig.6, the loss function value for the LSTM-Transformer model decreases rapidly during the initial iterations. By approximately 30 iterations, the loss function value reduces to the order of magnitude $10^{-4}$. At this point, the LSTM-Transformer model achieves relatively accurate predictions. In summary, the LSTM-Transformer model converges more rapidly than both the LSTM and GRU models. The learning performance of the three models on the training set and their predictive performance on the test set are illustrated in Fig.7 and Fig.8.

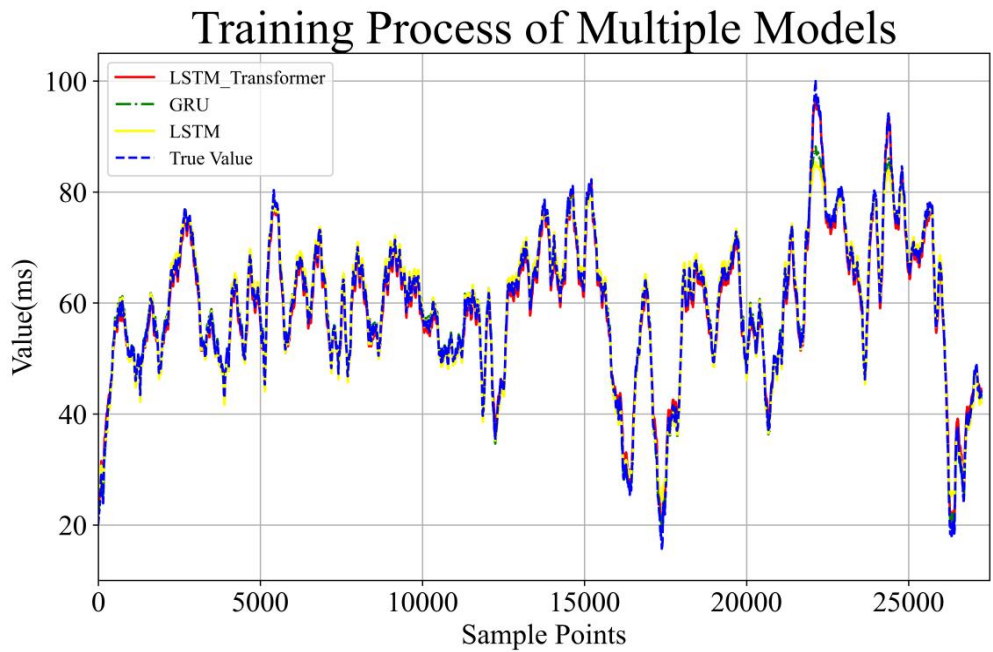

Fig.7 The learning curves of three models on the training set. The red line represents the learning process of the LSTM-Transformer Model on the training set, the green line represents the learning process of the GRU Model on the training set, the yellow line represents the learning process of the LSTM Model on the training set, and the blue line represents the actual dataset.

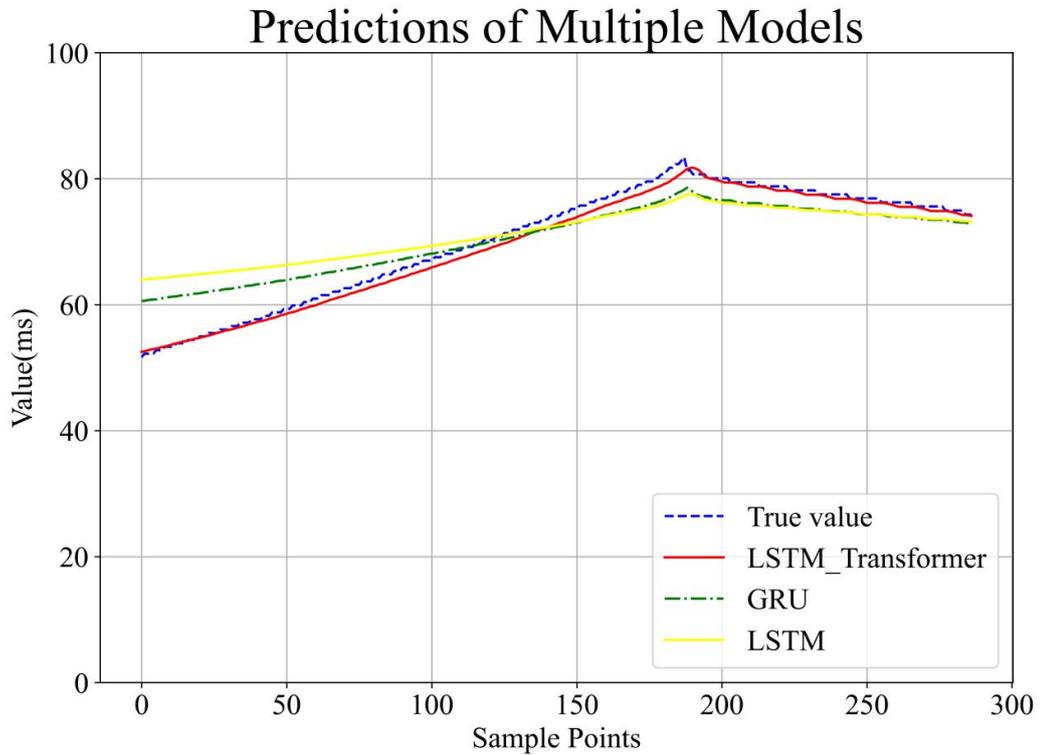

Fig.8 The learning curves of three models on the test set. The red line represents the predicted result of the LSTM-Transformer Model on the test set, the green line represents the predicted result of the GRU Model on the test set, the yellow line represents the predicted result of the LSTM Model on the test set, and the blue line represents the actual dataset.

## 3.2 Analysis of the Performance of Three Time Series Prediction Models:

**GRU Model:** The GRU Model exhibits excellent performance in feature extraction and prediction for data with relatively stable trends. However, a significant issue with the GRU Model is its relatively poor in predictive ability for data with pronounced fluctuations. Moreover, the GRU Model performs better than the LSTM Model on both the training and testing sets. When analyzing the model architecture, the GRU Model has fewer parameters compared to the LSTM Model, and facilitates better convergence during training. Furthermore, when comparing performances on the testing set, the LSTM-Transformer Model significantly outperforms the GRU Model. This suggests that the GRU Model exhibits a pronounced overfitting phenomenon, which is detrimental to its generalization in other application scenarios.

**LSTM Model:** Similar to the GRU Model, the LSTM Model performs well in feature extraction and prediction for data exhibiting relatively stable trends. However, for data with significant fluctuations, the LSTM model demonstrates a greater prediction error. The predictive performance of the LSTM model is inferior to that of its enhanced version, the LSTM-Transformer Model. Additionally, the predictive performance of the LSTM Model is slightly lower than that of the GRU Model.

**LSTM-Transformer Model:** The LSTM-Transformer Model outperforms the traditional LSTM Model in prediction accuracy across the board, with significantly better results on the test set than the GRU Model. Analyzing from the model's architectural perspective, the incorporation of the Attention layer from the Transformer significantly alleviates the long-term dependency issues present in the LSTM model by effectively weighting and computing a vast array of learned features, thus reducing prediction errors. Furthermore, the LSTM-Transformer model demonstrates solid performance on both the training and test sets, without showing signs of overfitting as observed with the GRU model. This indicates that the LSTM-Transformer model is well-suited for predicting future time data in pVAD experiments.

## 3.3 Comparison of the performance of MAE and MSE for three models

The quantification comparison of MAE and MSE for the three models on the training set is shown in Table.2:

Table.2 Comparison of Training Performance of the Three Models. In the training set, the prediction error of the LSTM-Transformer Model is significantly lower than that of both the GRU Model and LSTM Model.

| Models | MAE | MSE |
|---|---|---|
| GRU | 1.97 | 0.26 |
| LSTM | 2.05 | 0.29 |
| LSTM-Transformer | 0.79 | 0.14 |

The quantification comparison of MAE and MSE for the three models on the test set is shown in Table.3:

Table.3 Comparison of Predicting Performance of the Three Models. The LSTM-Transformer model once again demonstrates its significant superior performance. In terms of the MSE metric, the error of the LSTM-Transformer model is less than half of that of both the LSTM model and the GRU model.

| Models | MAE | MSE |
|---|---|---|
| GRU | 1.89 | 0.11 |
| LSTM | 1.93 | 0.13 |
| LSTM-Transformer | 0.71 | 0.05 |

## 3.4 Investigation of the robustness of the LSTM-Transformer Model in different dataset sizes

This study utilizes datasets of varying magnitudes to evaluate the robustness of the

LSTM-Transformer model across different scales of data, as in Table.4.

Table.4 Comparison of the Scalability Performance of the LSTM-Transformer Model

| Dataset size | Offline training time(s) | Online Fine-Tuning Time(s) | MSE |
| --- | --- | --- | --- |
| 500 | 15.6 | 0.7 | 0.14 |
| 1000 | 27.3 | 1.1 | 0.08 |
| 5000 | 119.8 | 1.8 | 0.05 |
| 10000 | 225.5 | 2.3 | 0.05 |

First, regarding computational time, the offline training process of the LSTM-Transformer Model does not exhibit explosive growth in computation time as the data volume increases. With a training data volume of up to 10,000 data points, the model's computation time is kept within 4 minutes. This proves that the LSTM-Transformer Model has a fast processing speed, demonstrating strong feasibility and application value. Additionally, it is found that the prediction error is relatively large when the data volume is at lower levels such as 500 or 1000. However, once the data volume reaches 5,000 or more, the model error converges to a stable value. This indicates that the LSTM-Transformer Model requires at least 5,000 data points as training support.

## 3.5 Investigation of the robustness of the LSTM-Transformer Model in different noise levels

Datasets with varying noise levels are applied to the LSTM-Transformer Model to investigate whether it can maintain robustness in the face of different noise levels, as in Table.5. The type of added noise is Gaussian noise. Gaussian noise is noise that follows a normal distribution. In many practical applications, measurement errors一such as sensor noise and image noise一can be approximately modeled as Gaussian distributions. This is due to the Central Limit Theorem, which states that the effects of multiple independent factors often lead the overall distribution of result to approach a normal distribution[16].

Table.5 Performance of the LSTM-Transformer Model Against Different Noise Levels

| Noise level (%) | LSTM-Transformer(MSE) | LSTM(MSE) |
| --- | --- | --- |
| 0 | 0.05 | 0.13 |
| 1 | 0.09 | 0.21 |
| 5 | 0.14 | 0.35 |
| 10 | 0.18 | 0.46 |

In clinical surgeries, the data collection process and other stages may encounter anomalies due to measurement errors, equipment failures, and human errors. The LSTM-Transformer Model maintains a low Mean Squared Error (MSE) when faced with datasets of varying noise levels, consistently keeping its MSE below half that of the LSTM Model. This demonstrates that the LSTM-Transformer Model exhibits high tolerance for abnormal data generated during data collection or other stages, indicating that the model possesses good generalizability and feasibility.

## 3.6 Investigation of the robustness of the LSTM-Transformer Model

## in hyperparameter sensitivity analysis

A hyperparameter tuning analysis of the LSTM-Transformer Model is conducted. By keeping other parameters constant (learning rate = 0.001, number of iterations = 100), various combinations of hyperparameters are experimented, as in Fig.9. Ultimately, it is determined that the optimal hyperparameter combination for the current dataset is: batch size = 32, dropout rate = 0.2.

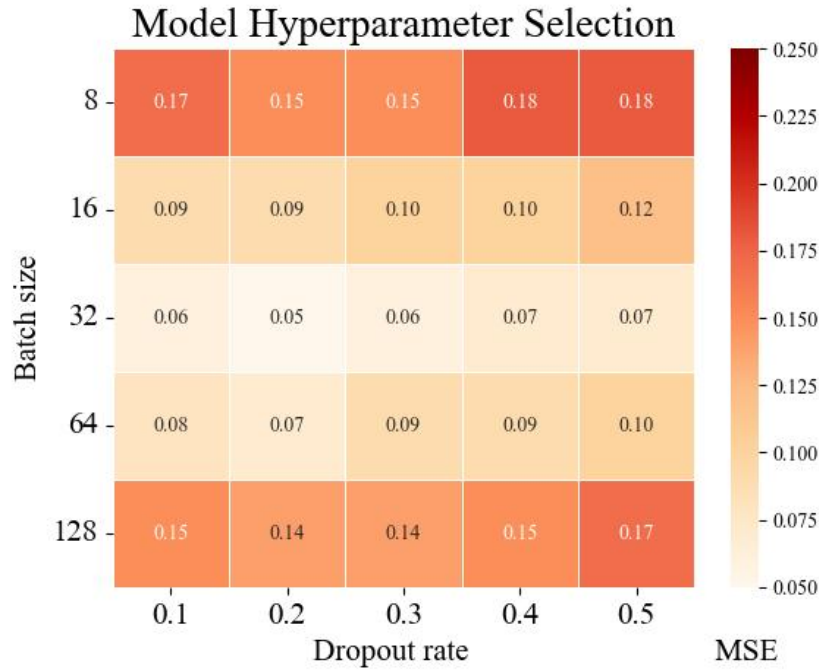

Fig.9 Comparison of the Model MSE under Different Hyperparameter Selections

Furthermore, from another perspective, it can be observed that, despite the selection of various hyperparameter combinations, the MSE of the LSTM-Transformer Model consistently remains below 0.2, indicating that the MSE is maintained at a relatively low level. This demonstrates the robustness of the LSTM-Transformer Model.

## 3.7 Investigation of the robustness of the LSTM-Transformer Model in model prediction time

The aforementioned prediction graphs for the test set illustrate approximately 5 minutes of forecasting. However, the model cannot provide accurate predictions for an infinite number of future time points; it can only guarantee accuracy within a fixed range of future time. The maximum time duration over which the model retains predictive accuracy needs to be identified. Therefore, the LSTM-Transformer Model is employed to predict data for time intervals of 5, 10, 15, 20, 25, and 30 minutes, and the MSE between these predictions and the actual data from animal experiments is calculated. The comparative results are shown in Table.6.

Table.6 Comparison of MSE for the three Models Predicting Future Time Intervals

| Models | 5min | 10min | 15min | 20min | 25min | 30min |
|---|---|---|---|---|---|---|
| GRU | 0.11 | 0.11 | 0.14 | 0.18 | 0.22 | 0.28 |
| LSTM | 0.13 | 0.13 | 0.17 | 0.21 | 0.26 | 0.33 |
| L-T Model | 0.05 | 0.05 | 0.05 | 0.06 | 0.09 | 0.14 |

From Table.6, it can be observed that the LSTM-Transformer model maintains a stable mean squared error (MSE) for the first 15 minutes. After the 15-minute mark, the MSE begins to increase. Therefore, the effective predictive time for the model is determined to be 15 minutes. This indicates that the LSTM-Transformer model can provide relatively accurate predictions for the next 15 minutes. For predictions extending beyond 15 minutes, the model requires online fine-tuning or retraining with new data. Previous analysis indicates that even in the case of comprehensive retraining, the required time is only 225.5 seconds; this scenario represents the worst-case for online fine-tuning. This worst-case scenario does not exceed 4 minutes; in comparison to the 15-minute prediction window, even with retraining, there remains a substantial time surplus throughout the prediction process. Therefore, the model is capable of delivering consistent and accurate predictions over extended future periods.

## 3.8 Hydraulic Experiment Validation of the AP-pVAD Model

The aforementioned NPQ Model and LSTM-Transformer Model are deployed to the motor controller of the pVAD, and three new hydraulic experiments are completed. The model deployment process is as follows: The computer (hardware environment: NVIDIA GeForce RTX 4060 Laptop GPU) receives real-time data transmitted by the sensors and calculates the model output results. The computer then transmits the model output results to the motor controller. The model of the motor controller (microcontroller) is STM32F103ZET6. The motor controller sets the corresponding speed adjustment instructions based on the received data.

First, the NPQ Model is validated. The desired flow rate is preset at 2.5 L/min, as this is the basic requirement for pVAD flow during high-risk PCI surgeries[8]. In this study, combinations of the desired pressure and flow are input into the NPQ model at different motor speeds to determine whether the pVAD can deliver results that are close to the desired pressure and flow values.

Three hydraulic experiments are conducted, and the comparison graphs of the experimental results and the desired flow rate and pressure are shown in Fig.10:

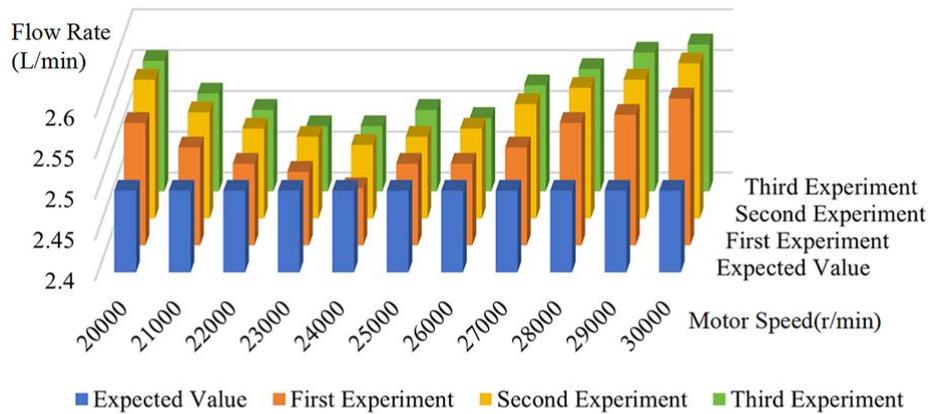

(a) Comparison of Hydraulic Experimental Results with Expected Values

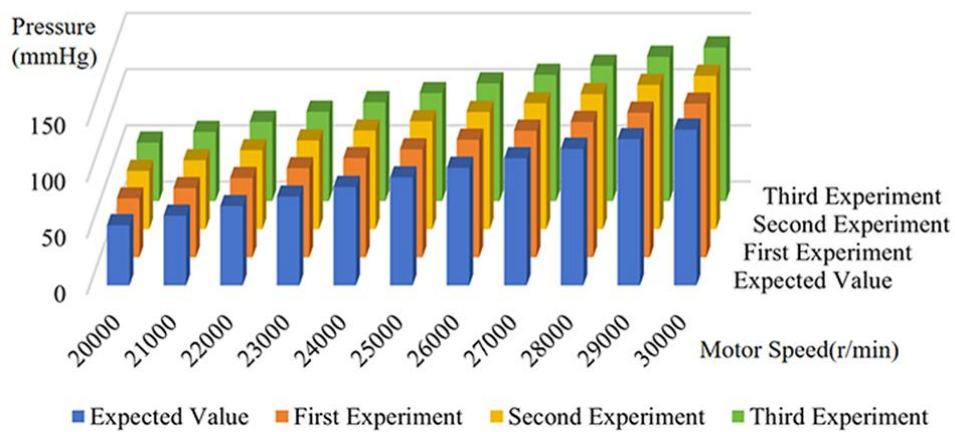

(b) Comparison of Hydraulic Experimental Results with Expected Values

Fig.10 Hydraulic Experiment Validation of the NPQ Model for Flow Rate and Pressure Provision. Figure(a) shows the comparison in flow rate, Figure(b) shows the comparison in pressure.

As shown in the figure(a), the flow rate remains approximately at 2.5 L/min across different rotational speeds, with a maximum absolute error of 0.09 L/min and a maximum relative error of 3.6%. Notably, the data within the range of 21,000 r/min to 28,000 r/min shows smaller errors, indicating that the NPQ Model provides the most accurate predictions within this rotational speed range. The average error is less than 0.05 L/min, demonstrating that the NPQ Model can enable the pVAD to deliver relatively accurate flow rates. As shown in the figure(b), under the condition of a fixed flow rate of 2.5 L/min, the pressure increases with the rise in rotational speed. The maximum absolute pressure error occurs at a rotational speed of 30,000 r/min, measuring 2.15 mmHg, with a maximum relative error of 1.5%. This indicates that the NPQ Model enables the pVAD to provide relatively accurate pressure .

The LSTM-Transformer Model's predictive performance for characteristic pulsation time points is validated on a pulsatile hydraulic test platform. By adjusting the pulsation frequency of the hydraulic test platform motor, whose function is to generate pulsating water flow, a set of true values (expected values) for the pulsation time characteristic points is obtained. This data is then input into the LSTM-Transformer Model to compare whether the predicted values output by the LSTM-Transformer Model are close to the expected values, as in Fig.11.

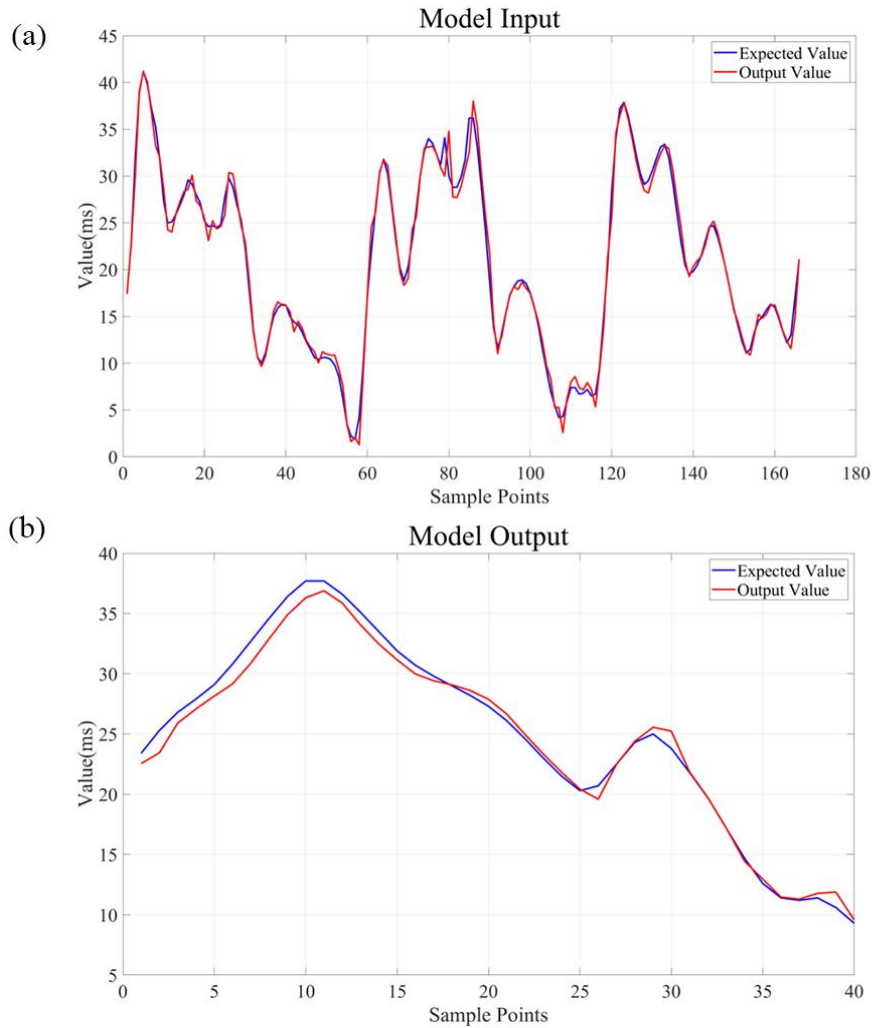

Fig.11 Validation of the LSTM-Transformer Model's performance on the training set in Fig(a) and test set in Fig(b). The blue line represents the true values of the pulsating hydraulic experiment dataset, while the red line indicates the training performance and the predictions of the LSTM-Transformer Model on the pulsating hydraulic experiment dataset.

As illustrated in Fig.11, the predicted results outputted by the model closely approximate the expected (true) values. The maximum error between the model's output and the true values does not exceed 2 ms. This indicates that the LSTM-Transformer Model can predict the pulsatile timing characteristics with considerable accuracy, allowing the pVAD to provide blood flow perfusion closely synchronized with the expected values.

The aforementioned dataset is applied to the LSTM-Transformer Model, GRU Model and LSTM Model, with multiple predictions being executed for each. Subsequently, a statistical analysis is conducted to compare the Maximum Error across all models, as in Table.7.

Table.7 Comparison of Maximum Error across three models

| Models | Maximum Prediction Error(ms) |
|---|---|
| GRU | 4.37 |
| LSTM | 5.96 |
| LSTM-Transformer | 1.78 |

The maximum prediction error of the LSTM-Transformer Model on the aforementioned dataset is significantly smaller than that of the GRU Model and the LSTM Model, demonstrating that the LSTM-Transformer Model represents a notable improvement in prediction performance compared to some traditional deep learning prediction models.

## 3.9 Animal Experiment Validation of the AP-pVAD Model

The aforementioned NPQ Model and LSTM-Transformer Model were deployed to the motor controller of the pVAD, and one new animal experiment is completed. The deployment process for the animal experiment model is consistent with that of the hydraulic experiment model described above.

The following figure shows the variation of ascending aortic pressure over time during the pVAD startup process (the motor speed gradually increases uniformly from 18,000 r/min to 25,000 r/min, with the pVAD starting at a speed of 18,000 r/min). The data collection method involved attaching a pressure sensor to the rear guide vane, sampling at a frequency of 150 Hz. The peak value (SBP) and trough value (DBP) of the AOP in sheep are recorded, and the mean arterial pressure(MAP) is calculated. The following graph is plotted with motor speed on the x-axis, as in Fig.12.

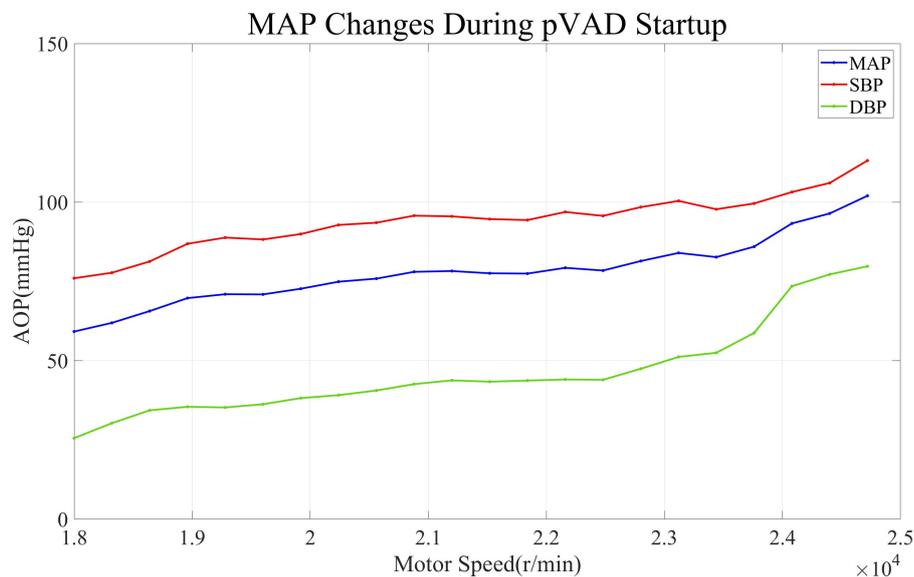

Fig.12 Changes in MAP, SBP and DBP During the pVAD Startup Process

As shown in Fig.12, the MAP of the sheep increased from 59.15 mmHg before the pVAD startup to 101.97 mmHg after the startup. Reference[41] indicates that perioperative hypotension is closely related to brain injury and renal impairment. The pVAD startup significantly improved the blood pressure of the sheep. The pVAD is implanted in the sheep and operated continuously.

During the 27 hours after the surgeon closed the chest and left the operating table, heart rate and blood oxygen levels are recorded every hour. The results are shown in Fig.13:

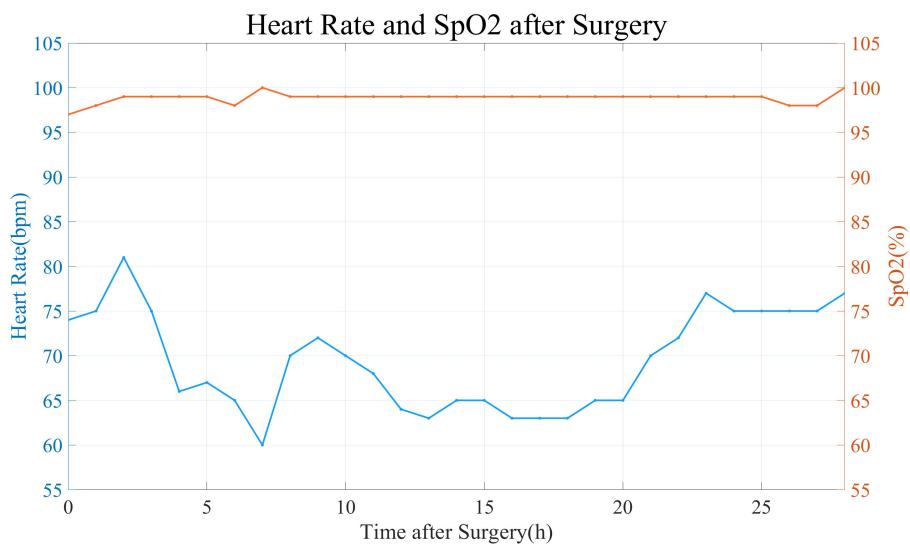

Fig.13 Changes in heart rate and blood oxygenation assisted by pVAD

As shown in Fig.13, after the pVAD assistance, the heart rate of the sheep decreases due to the increase in cardiac output. The sheep's heart rate remains within a stable range (60 bpm - 81 bpm), and blood oxygen levels stabilized around normal values. This indicates that the pVAD effectively maintains the normal vital signs of the sheep. The sheep survives normally for over 27 hours with pVAD assistance.

## 4. Discussion and Conclusion

In summary, a model of the artificial pulsation for a pVAD is proposed (AP-pVAD model). AP-pVAD model consists of two parts: NPQ Model and LSTM-Transformer Model . The model calculates the motor speed-blood pump pressure-blood pump flow mathematical model (NPQ Model) by derivation of Bernoulli equation, based on hydraulic experiment data from the pVAD. Clinical pressure and flow requirements are input into the model, then the NPQ Model provides the expected motor speed for the given conditions. Subsequently, the Attention module of Transformer neural network is integrated into the LSTM neural network to form the new LSTM-Transformer Model, which is later trained using an animal experimental dataset to predict the pulsation time characteristic points for adjusting the motor speed. This allows the pVAD to achieve pulsatile blood flow that is synchronized both numerically and temporally with native human hemodynamics.

The NPQ Model indicates that the pVAD motor speed has a linear relationship with pressure and a quadratic relationship with flow. The LSTM-Transformer Model demonstrates that the LSTM-Transformer Model is a better method for predicting the pulsation time characteristic points of pVAD. Compared to traditional time series forecasting methods like GRU and LSTM, the LSTM-Transformer Model converges faster and has smaller prediction errors. Additionally, in model robustness research, the performance of the LSTM-Transformer Model is evaluated across various conditions, including different dataset sizes, different noise levels and hyperparameter sensitivity. LSTM-Transformer model demonstrates minimal prediction error and better robust performance under conditions of limited dataset sizes, elevated noise levels, and diverse hyperparameter combinations, proving to be an effective approach.

The AP-pVAD Model is validated in three hydraulic experiments and an animal experiment to verify its feasibility and effectiveness. The pVAD provides a pulsatile flow that matches the expected values. In the three hydraulic experiments, the pressure provided by pVAD, calculated using the NPQ Model, has a maximum error of only 2.15 mmHg compared to the expected values. The pulsation time characteristic points predicted by the LSTM-Transformer Model shows a maximum prediction error of 1.78ms, while GRU Model shows 4.37ms and LSTM Model shows 5.96ms. The in-vivo test of pVAD in animal experiment has significant improvements in parameters such as aortic pressure. Animals survive for over 27 hours after the initiation of pVAD operation.

The AP-pVAD Model exhibits good robustness and wide applicability. The pVAD itself, along with high-risk PCI and cardiogenic shock surgeries, incorporates many measurable continuous time series data. The LSTM-Transformer Model is not stringent regarding time series forecast data volume and noise. Therefore, the LSTM-Transformer Model has shown adaptation of time series forecast data in various clinical scenarios in the future. This provides valuable reference points for future studies on various LVAD.

There is room for improvement in this work. For instance, the pulsation time characteristic points only selected the time point at the beginning of the AOP rise, which is chosen for computational efficiency and simplicity. Future research could consider multiple characteristic points on the AOP curve within a single cardiac cycle as pulsation time characteristic points. Changing the model input from a single point to multiple points will inevitably increase computation time; however, advancements in computational hardware will alleviate this issue.

In future studies, these deficiencies will be addressed, and more hydraulic and animal experimental data will be applied to the AP-pVAD Model to further reduce model errors and verify the conclusions drawn in this research.

## 5. Statement

All animal experiments in this study comply with the ARRIVE (Animal Research: Reporting of In Vivo Experiments) guidelines. All animal experiments in this study are carried out in accordance with the Guidance on the operation of the Animals (Scientific Procedures) Act 1986 and associated guidelines, EU Directive 2010/63 for the protection of animals used for scientific purposes, the NIH (National Research Council) Guide for the Care and Use of Laboratory Animals (PDF), or those of an equivalent internationally recognized organization. This study affirms that the guidelines listed above have been adhered to.

During the preparation of this work the authors used ChatGPT in order to improve English readability. After using this tool, the authors reviewed and edited the content as needed and take full responsibility for the content of the publication.

# 6. Acknowledgements

This work was supported by the National Key R&D Program of China(Grant No. 2022YFC2402600) and National Natural Science Foundation of China (Grant Nos. 12388101, 12072174).

# 7. Conflict of Interest

The authors declare that the research was conducted in the absence of any commercial or financial relationships that could be construed as a potential conflict of interest.